# A Josephson relation for fractionally charged anyons.

M. Kapfer[1], P. Roulleau[1], M. Santin[1], I. Farrer[2], D. A. Ritchie[3], and D. C. Glattli[1*]

**Affiliations:**

[1]Service de Physique de l'Etat Condensé, IRAMIS/DSM (CNRS UMR 3680), CEA Saclay, F-91191 Gif-sur-Yvette, France..

[2]Department of Electronic and Electrical Engineering, University of Sheffield, Mappin Street, S1 3JD, UK.

[3]Cavendish Laboratory, University of Cambridge, J.J. Thomson Avenue, Cambridge CB3 0HE, UK.

*Correspondence to: christian.glattli@cea.fr

**Abstract:** Anyons occur in two-dimensional electron systems as excitations with fractional charge in the topologically ordered states of the Fractional Quantum Hall Effect (FQHE). Their dynamics are of utmost importance for topological quantum phases and possible decoherence free quantum information approaches, but observing these dynamics experimentally is challenging. Here we report on a dynamical property of anyons: the long predicted Josephson relation $f_J=e^*V/h$ for charges $e^*=e/3$ and $e/5$, where e is the charge of the electron and h is Planck's constant. The relation manifests itself as marked signatures in the dependence of Photo Assisted Shot Noise (PASN) on voltage V when irradiating contacts at microwaves frequency $f_J$. The validation of FQHE PASN models indicates a path towards realizing time-resolved anyon sources based on levitons.

**Main Text:** The Quantum Hall Effect (QHE) occurs in two-dimensional electron systems (2DES) when strong magnetic fields quantize the electron cyclotron energy into Landau levels. For integer Landau level filling factor $\nu=p$, the Integer QHE (IQHE) shows a topologically protected quantized Hall conductance $pe^2/h$ with zero longitudinal conductance (*1*). For very low disorder samples, the Coulomb repulsion favors topologically ordered phases at rational $\nu=p/q$ showing a Fractional QHE (FQHE) with fractional Hall (*2*) and zero longitudinal conductance. For electrons filling the first Landau Level ($\nu<1$), the states with $\nu=1/(2k+1)$, where k is an integer, are well described by Laughlin states (*3*). The elementary excitations, or quasiparticles, bear a fraction $e^*=e/(2k+1)$ of the elementary charge e (*3-6*) and are believed to obey a fractional anyonic (*7*) statistics intermediate between bosons and fermions. For $\nu<1$, the Jain states (*8*) with $\nu=p/(2kp+1)$, p and k integer, display $e^*=e/(2kp+1)$ fractionally charged excitations (*9*); these excitations are composite fermions, i.e., electrons to which $-2k$ flux quanta $\phi_0=h/e$ are attached (here h is Planck's constant). For higher Landau Level filling, even-denominator FQHE states are found, such as the 5/2 state that hosts Majorana excitations and $e^*=e/4$ non-abelian anyonic quasiparticles (*10,11*); these have possible applications to topologically protected quantum computation. An important breakthrough in this context would be the time domain manipulation of anyons allowing braiding interference using Hong Ou Mandel correlations (*12-14*); understanding the dynamics of anyons is thus of utmost importance.

Central to this understanding is achieving an experimental observation of the Josephson relation $f_J = e^*V/h$, which implies that $e^*$ anyons are elementary excitations that undergo photon assisted energy transitions while a voltage V is applied on the conductor. The Josephson relation has a long history starting from the discovery of the AC Josephson Effects (*21,22*) in superconductors. When two tunnel coupled superconductors are biased by a voltage $V_{dc}$, a steady current oscillation occurs at frequency $f_J$ providing evidence of the Cooper pair charge e*=2e. The inverse AC Josephson effect occurs when a superconducting junction is irradiated at frequency $f$; when the bias voltage Josephson frequency $f_J$ matches a multiple of $f$, photon-assisted singularities called Shapiro steps appear in the I-V characteristics (*22*). The AC Josephson effects arise from the quantum beating between tunnel coupled Cooper pair condensates at energies separated by $eV_{dc}$. For normal metals, described by a Fermi sea, no steady Josephson oscillations are expected but transient current oscillations at frequency $f_J=eV_{dc}/h$ were demonstrated in numerical quantum simulations for two voltage shifted Fermi seas put in quantum superposition in an electronic interferometer (*23*). Also in normal metals, the Josephson frequency manifests itself in the high frequency shot noise when the bias voltage equals the emission noise frequency. Reciprocally, the low frequency photo-assisted shot noise (PASN) shows Josephson like singularities when microwaves irradiate a contact at frequency $f=f_J$.

First predicted (*24*) for mesoscopic conductors, PASN is also expected to occur in interacting electronic systems, like the FQHE (*17-20*). In the absence of microwave irradiation the (DC) shot noise is the result of the quantum beating of two voltage shifted Fermi seas when scattering in the conductor mixes the carrier states. For a single mode normal conductor (e*=e) with conductance $g_0=e^2/h$ and a unique scatterer of reflection probability R, the zero temperature current noise spectral density under DC bias $V_{dc}$ is given by $S_I^{DC} = 2g_0 R(1-R)e|V_{dc}|$. When adding an AC voltage to the biased contact: $V(t)=V_{dc} + V_{ac}(t)$, with $V_{ac}(t)=V_{ac}\cos(2\pi ft)$, the phase of all carriers emitted by the contact gets a time dependent part $\phi(t) = \frac{e*}{h}\int_{-\infty}^{t} V_{ac}(t')dt'$ giving rise to energy scattering. The emitted carriers end in a superposition of quantum states with energy shifted by $lhf$ and probability amplitude $p_l = \frac{1}{T}\int_0^T e^{-i\phi(t)} e^{i2\pi l f t} dt$, where $l$ is integer and $T=1/f$.

Using the voltage $V_{dc}$ in units of the Josephson frequency $f_J = e^* V_{dc}/h$, the predicted PASN spectral density can be written as:

$$S_I^{PASN}(f_J) = \sum_{l=-\infty}^{l=\infty} |p_l|^2 S_I^{DC}(f_J - lf) \qquad (1)$$

where $S_I^{DC}$ is the DC shot noise measured when $V_{ac}=0$. Equation 1 expresses that the measured observable is the result of the sum of simultaneous measurements with shifted voltage $V_{dc} \rightarrow V_{dc} + lhf/e^*$ and weighted by the probability $|p_l|^2$, as the Fermi sea of the driven contact is in a quantum superposition of states with energy shifted by $lhf$. Interestingly the zero bias voltage DC shot noise singularity, $\sim|V_{dc}|\sim|f_J|$, is replicated whenever $f_J = e^*V_{dc}/h = lf$, indicating the Josephson frequency and hence the existence of photon assisted transitions by anyons of charge $e^*$. This effect parallels the Shapiro steps of superconducting junctions IVC, i.e. the inverse AC Josephson effect (*22*). The PASN singularity for $f_J = eV_{dc}/h = f$ has been observed in normal conductors ($e^*=e$), such as diffusive metallic wires (*25*), Quantum Point Contacts (*26*), and tunnel junctions (*27*). For interacting systems, Eq. 1 has been derived in (*29*) and observed in (*28*) for superconducting/normal junctions ($e^*=2e$). In FQHE systems, the concept of fractional Josephson frequency was introduced in Refs. (15) and (16), which discussed photo-assisted processes (*15*) or finite frequency noise (*16*). The concept and the terminology was later used in FQHE PASN models (*17*). Equation 1 was explicitly shown in (*18*) and is implicit in PASN expressions of refs (*17-20*). However, experimentally combining high magnetic fields, sub-fA/Hz$^{1/2}$ current noise and >10 GHz microwaves at ultra-low temperature (~20mK) is highly challenging.

In this work, we combine microwave frequency irradiation with low frequency shot noise measurements to provide evidence of the above Josephson relation and a conclusive test of PASN models.

A schematic view of the set-up and the sample is shown in Fig.1A. In topologically insulating QHE conductors the current flows along p chiral edge modes for filling factor $\nu=p/(2p+1)$. To inject current or apply a microwave excitation, metallic contacts connect the edges to an external circuit. A narrow tuneable constriction called Quantum Point Contact (QPC) is formed by applying a negative voltage $V_G$ to split gates to induce backscattering by the quantum mixing of counter-propagating edge modes. Carriers incoming from contact 0 and scattered by the QPC contribute to transmitted and backscattered currents, $I_t$ and $I_B$ respectively. These currents are measured at contact (1,2) via the voltages $V_{1(2)}=(h/e^2\nu_B)I_{t(B)}$ where $\nu_B$ is the filling factor in the lead (far from the QPC). The partitioning of the carrier generates a current noise $S_I$ which is measured by recording the negative cross-correlation of the voltage fluctuations $\Delta V_{1(2)}=(h/e^2\nu_B)\Delta I_{t(B)}$ giving $S_I = -\langle \Delta I_t \Delta I_B \rangle / \Delta f$. $\Delta f$ is the bandwidth of the low frequency resonant detecting circuit as described in *(30)*.

We focus on bulk filling factor $\nu_B=2/5$, which conveniently allows us to probe both e/3 and e/5 anyons. The two co-propagating chiral edge modes of the 2/5 Jain state are revealed by sweeping the QPC gate voltage $V_G$ (Fig.1B). Starting with a $(2/5)e^2/h$ conductance plateau we observe a second conductance plateau $(1/3)e^2/h$ at lower $V_G$. This corresponds to a fully reflected inner channel with conductance $g_2=(2/5-1/3)e^2/h$ whereas the outer edge channel with conductance $g_1=(1/3)e^2/h$ is fully transmitted. To probe the e/3 charged excitations of the 1/3-FQHE state locally formed at the QPC, $V_G$ is set to -0.090V (point (A) in Fig.1B) so as to induce a weak backscattering (WB) between counter-propagating outer edge modes with reflection probability R=0.026 . Next, we apply a dc voltage $V_{dc}$ to the injecting contact. The incoming current of the outer edge mode $I_0=(1/3)(e^2/h)V_{dc}$ divides into a backscattering current $I_B \approx RI_0$ and a forward current $I=I_0-I_B$. In the FQHE, the chiral modes form chiral Luttinger liquids (*14*). At finite backscattering transport becomes energy dependent, giving non-linear variations of $I_B$ with voltage $V_{dc}$. A complete modelling is difficult and comparison to experiments is only easy in the WB regime (R<<1). The small backscattered current $I_B(V_{dc})$ results from rare quasiparticle tunnelling events following Poissonian statistics. In this limit, the DC shot noise cross-correlation is (*16,31,32*):

$$S_I^{dc}(V_{dc}) = 2e^* I_B(V_{dc})[\coth(e^*V/2k_BT_e) - 2k_BT_e/e^*V_{dc}] \qquad (2)$$

with $e^*=e/3$ for the 1/3-FQHE regime considered here and $T_e$ the electronic temperature.

Figure 2B, black dots, shows DC shot noise data. The black dashed line, computed using Eq. 2 compares well with the data for $e^*=e/3$. Here a constant R≈0.026 versus $V_{dc}$ is used as $I_B(V_{dc})$ is found to be almost linear (Fig.S3).

Next, to show the Josephson relation using PASN, the AC voltage $V_{ac}(t)=V_{ac}\cos2\pi ft$ is superimposed on $V_{dc}$ with f=22GHz. The blue and red dots show the measured (PASN) noise for several $V_{ac}$ corresponding to -61 and -67dBm nominal RF power (disregarding rf lines losses) sent to the contact. At low $V_{dc}$, the PASN noise increases with power; for $V_{dc}$ above ≈250µV, the PASN noise merges into the DC shot noise curve The change in the slope of the noise variation at this characteristic voltage is suggestive, but not conclusive of the expected PASN noise singularity. In order to reveal pure photon-assisted contributions to PASN, guided by the form of Eq. 1, we cancel the *l*=0 term by subtracting the independently measured DC shot noise data from the raw PASN data. This defines the Excess PASN $\Delta S_I = S_I^{PASN}(V_{dc}) - |p_0|^2 S_I^{dc}(V_{dc})$. Finding the condition to cancel the (*l*=0) DC shot noise term in $\Delta S_I$ provides the value of $|p_0|^2=J_0(\alpha)^2$ for the excitation $V_{ac}=\alpha hf/e^*$ used. This is done for three RF powers: 67, 63, and 61dBm (Fig.S4). For clarity and better data statistics, only the calculated average of the three excess PASN curves is shown in Fig.2C, blue dots. Neglecting $|l|>1$ photon process and using the average $|p_1|^2$ values obtained from the average $|p_0|^2$ value: $<|p_1|^2>=(1-<|p_0|^2>)/2$ the theoretical Excess PASN is:

$$\Delta S_I(f_J{}^*)= \left\langle |p_1|^2 \right\rangle S_I{}^{DC}(f_J - f) + \left\langle |p_1|^2 \right\rangle S_I{}^{DC}(f_J + f) \qquad (3)$$

Equation 3 is plotted using $f_J=(e/3)V_{dc}/h$ as a blue dashed line in Fig.2C. The extracted value of $<|p_1|^2>$ and the choice of $f_J$ account well for the measured Excess PASN variation, strongly supporting the validity of Eq. 1.

A further validation of Eq. 1 is given by changing the excitation frequency. We have repeated similar measurements and analyses for f=17GHz and f=10GHz. In Fig.2C, the green and red dot curves show excess PASN data and the green and red dashed line provide convincing comparisons to Eq.(3) using the average parameter $<|p_1|^2>$ extracted from each experimental curve and fixing $e^*=e/3$ when calculating $f_J$. Using now $e^*$ as a free parameter the quantity $V_J= hf/e^*$ (which signals the onset of excess PASN) is extracted from best fit of Eq. 3 to the excess PASN data for each frequency f. When $V_J$ is plotted in Josephson frequency units $(e/3)V_J/h$ versus f (Fig.2D), a linear fit to the data gives $e^*=e/(3.06 \pm 0.20)$, yielding the fractional charge of the anyon. For comparison, the red dashed straight line, slope one, corresponds to $e^*=e/3$ exactly.

We then confirmed that we are measuring the Josephson frequency by changing the excitation charge. We consider the WB regime of the inner edge of the 2/5 FQHE Jain state, whose nominal conductance is $(2/5-1/3)e^2/h$ (Fig.3A). The backscattered current is now $I_B=R(1/15)e^2/hV_{dc}$. We set $V_G$ to -0.03V for which R=0.064, point (B) of Fig.1B. Figure 3B shows the DC noise data (black dots). A comparison of data to Eq. 2 with $e^*=e/5$ and R=0.064 (black dashed lines) confirms a one-fifth quasiparticle charge (*9*). The PASN for various RF power at 17 GHz is shown as coloured circles. As previously done for e/3 charges, we average the PASN data at 17GHz for three different Rf power (-60, -58 and -55dBm). The resulting mean excess noise $\langle \Delta S_I \rangle = \langle S_I^{PASN}(V_{dc}) \rangle - \langle |p_O|^2 \rangle S_I^{dc}(V_{dc})$ is plotted in Fig.3C (blue dots). The data compare well with Eq. 3 (red dashed line), using $f_J=(e/5)V_{dc}/h$ and $\langle|p_1|^2\rangle=(1-\langle|p_0|^2\rangle)/2$, including the finite temperature $T_e$=30mK in the DC Shot Noise, Eq. 2. Note that we have no independent way to determine $T_e$. A similar procedure is done for 10GHz excitation. Then, the voltage $V_J$=hf/e* characterising the onset of excess PASN is left as a free parameter to fit the excess PASN data and is plotted in units of Josephson frequency $f_J=(e/5)V_J/h$ versus f, in Fig.3D. The dashed line, slope one, corresponds to e*=e/5 exactly. A linear fit of the actual $V_J$ versus f passing by zero gives e*=e/(5.17±0.31).

To measure $f_J$ for non-abelian anyons at $\nu_B$=5/2 will require ultra-high mobility samples. A possible route is the realization of a single anyon source based on levitons (*12*) using periodic Lorentzian voltage pulses instead of a sine wave. The PASN caused by periodic levitons is also given by Eq. 1 except that all the $p_l$ for l<0 vanish, characterizing a minimal excitation state (*33*) without hole-like excitations, see (*30*) for more details. A charge e Leviton sent to a QPC in the WB regime would provide a convenient time controlled single anyon source with Poisson's statistics (*13,20*). Combining two similar sources opens the way for anyon braiding interference through Hong Ou Mandel tests of anyonic statistics (*13*) (Fig.S8).

**References and Notes:**

1. K. v. Klitzing, G. Dorda, and M. Pepper, New Method for High-Accuracy Determination of the Fine-Structure Constant Based on Quantized Hall Resistance. *Phys. Rev. Lett.* **45**, 494-497 (1980).

2. D. C. Tsui, H. L. Stormer, and A. C. Gossard, Two-Dimensional Magnetotransport in the Extreme Quantum Limit, *Phys. Rev. Lett.* **48**, 1559-1562 (1982)

3. R. B. Laughlin, Anomalous Quantum Hall Effect: An Incompressible Quantum Fluid with Fractionally Charged Excitations. *Phys. Rev. Lett.* **50**, 1395-1398 (1983)

4. L. Saminadayar, D. C. Glattli, Y. Jin, and B. Etienne, Observation of the e/3 Fractionally Charged Laughlin Quasiparticle. *Phys. Rev. Lett.* **79**, 2526-2529 (1997)

5. R. de-Picciotto, M. Reznikov, M. Heiblum, V. Umansky, G. Bunin & D. Mahalu. Direct observation of a fractional charge. *Nature* **389**, 162-164 (1997)

6. M. Hashisaka, T. Ota, K. Muraki, and T. Fujisawa, Shot-Noise Evidence of Fractional Quasiparticle Creation in a Local Fractional Quantum Hall State. *Phys. Rev. Lett.* **114**, 056802 (2015).

7. D. Arovas, J. R. Schrieffer, and Frank Wilczek, Fractional Statistics and the Quantum Hall Effect. *Phys. Rev. Lett.* **53**, 722-723 (1984)

8. J. K. Jain Composite-fermion approach for the fractional quantum Hall effect. *Phys. Rev. Lett.* **63**, 199-202 (1989)

9. M. Reznikov, R. de Picciotto, T. G. Griffiths, M. Heiblum and V. Umansky Observation of quasiparticles with one-fifth of an electron's charge. *Nature* 3**99**, 238–241 (1999)

**Acknowledgments:** We thank P. Jacques for technical help, P. Pari, P. Forget and M. de Combarieu for cryogenic support, and we acknowledge discussions with M. Freedman, I. Safi, Th. Martin, X. Waintal, H. Saleur and members of the Saclay Nanoelectronics Group. DR and IF acknowledge the EPSRC; **Funding:** the ANR FullyQuantum AAP CE30 grant is acknowledged; **Author contributions:** D.C.G. designed the project. M.K. made the measurements and with P.R. and D.C.G. did the data analysis. M. Santin has contributed to early experimental set-up and not to the present measurements. All authors, except M. Santin, discussed the results and contribute to writing the article. Samples were nanolithographied by M.K. on wafers from D.R. and I.F;

**Supplementary Materials:**

Materials and Methods

Supplementary text

Figures S1-S8

Data for figures 2, 3, S2-S7 available under DOI: 10.5281/zenodo.2388813

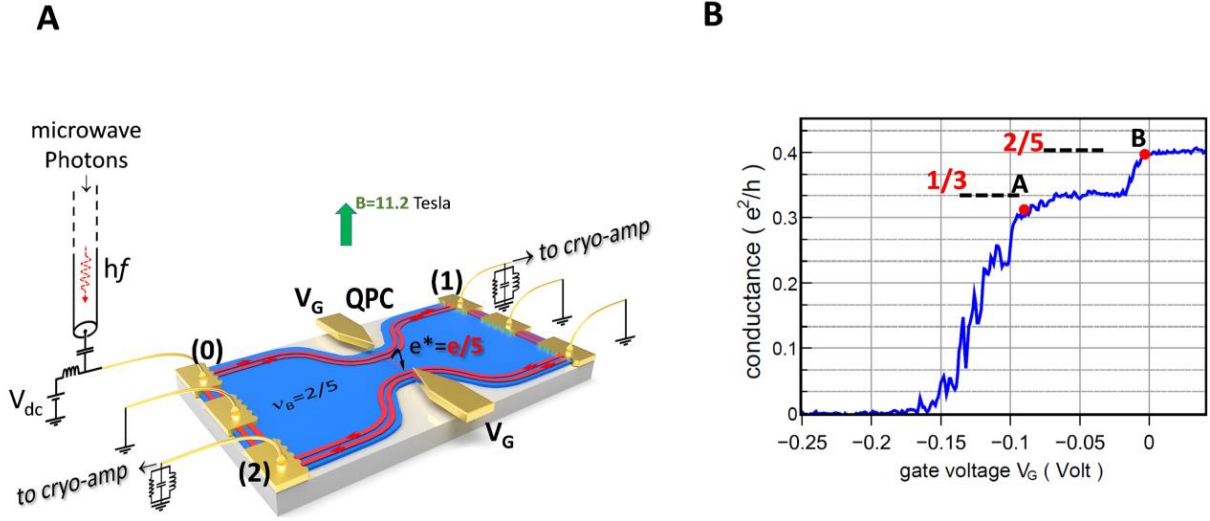

**Fig. 1. Schematics for PASN measurements.** (**A**) a DC voltage $V_{dc}$ applied to contact (0) injects carriers at bulk filling factor $\nu_B=2/5$ into two chiral fractional edge modes (red lines). The carriers are partitioned by a quantum point contact (QPC) into transmitted and reflected current, which are absorbed at the grounded contacts giving rise to voltages $V_1 = I_t h/\nu_B e^2$ and $V_2 = I_B h/\nu_B e^2$ at contacts (1) and (2), respectively. The negative voltage fluctuation cross-correlation $\Delta V_1 \Delta V_2 = -(h/\nu_B e^2)^2 S_I \Delta f$ is recorded to obtain the noise $S_I$. The voltages are sent to two identical resonant circuits followed by cryogenic amplification and fast digital acquisition. A computer performs the FFT cross-correlation. The RF excitation from a microwave photon source is added to $V_{dc}$ and sent to contact (0). (**B**) QPC conductance $G_t = dI_t/dV_{dc}$ at bulk filling factor $\nu_B=2/5$ versus the QPC gate voltage $V_G$. The $(2/5)e^2/h$ plateau is followed by a $(1/3)e^2/h$ plateau signaling complete reflection of the inner 2/5 fractional edge channel. Points (A) and (B) show the weak backscattering conditions for measurements with fractional carriers e/3 and e/5 respectively.

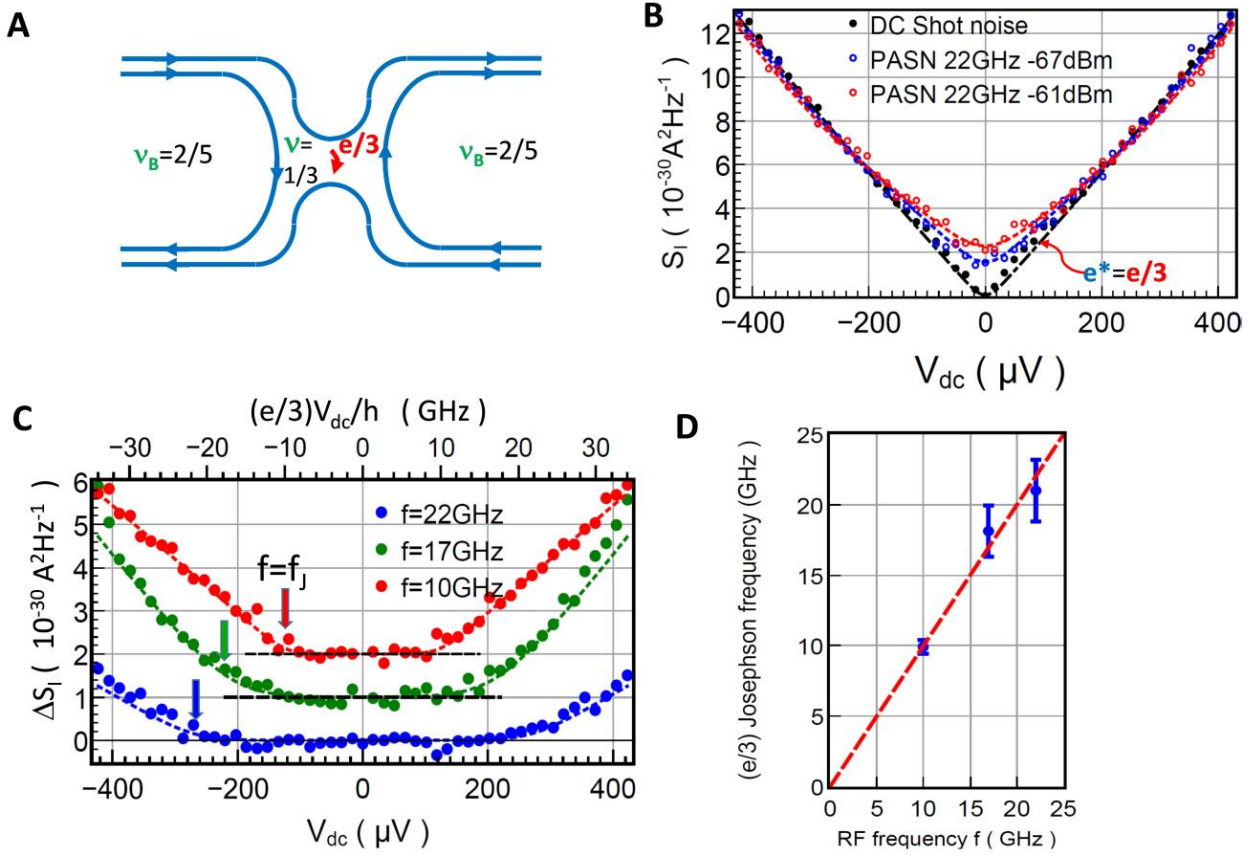

**Fig. 2. Josephson relation for 1/3 FQHE state.** (**A**) The fully reflected 2/5 inner edge state gives rise to a ν=1/3 FQHE state at the QPC. For $V_G$=-0.090V, point (A) of Fig.1B, the counter-propagating outer edges states are weakly coupled, allowing to probe e/3 backscattered carriers. (**B**) Raw shot noise measurements: Black dots show the DC shot noise (i.e. with only $V_{dc}$ applied at contact (0) and no RF). The dashed line is Eq. 2 with e*=e/3 and constant R=0.026 (point (A) of Fig.1A). Blue and red open circles are noise measurements for 22GHz -67dBm and -61dBm RF irradiation. Blue and red dashed line curves plot Eq. 1 with $f_J$=(e/3)$V_{dc}$/h and using $|p_0|^2$ and $|p_1|^2$ deduced from the analysis of Fig.2C. (**C**) Excess PASN $\Delta S_I$ (blue, green and red dots) for three frequencies 22, 17 and 10GHz respectively. The average of measurements at several excitation powers is shown to improve the noise statistics. The blue, green and red dashed lines, computed from Eq. 3 using $f_J$=(e/3)$V_{dc}$/h, compare well to the data. For clarity the constant $\Delta S_I(V_{dc}=0)$ has been subtracted from the excess PASN and the 17 and 10 GHz data have been offset. **(D)** Determination of carrier charge from the Josephson relation. A best fit of $\Delta S_I$ gives, for each frequency, the threshold voltages $V_J = hf/e^*$ above which $\Delta S_I$ rises. They are plotted in units of $(e/3)V_J/h$ versus f (blue points with s.e.m. error bars); a linear fit gives $e^* = e/(3.06 \pm 0.20)$. For comparison, the red dashed line corresponds to $e^* = e/3$ exactly.

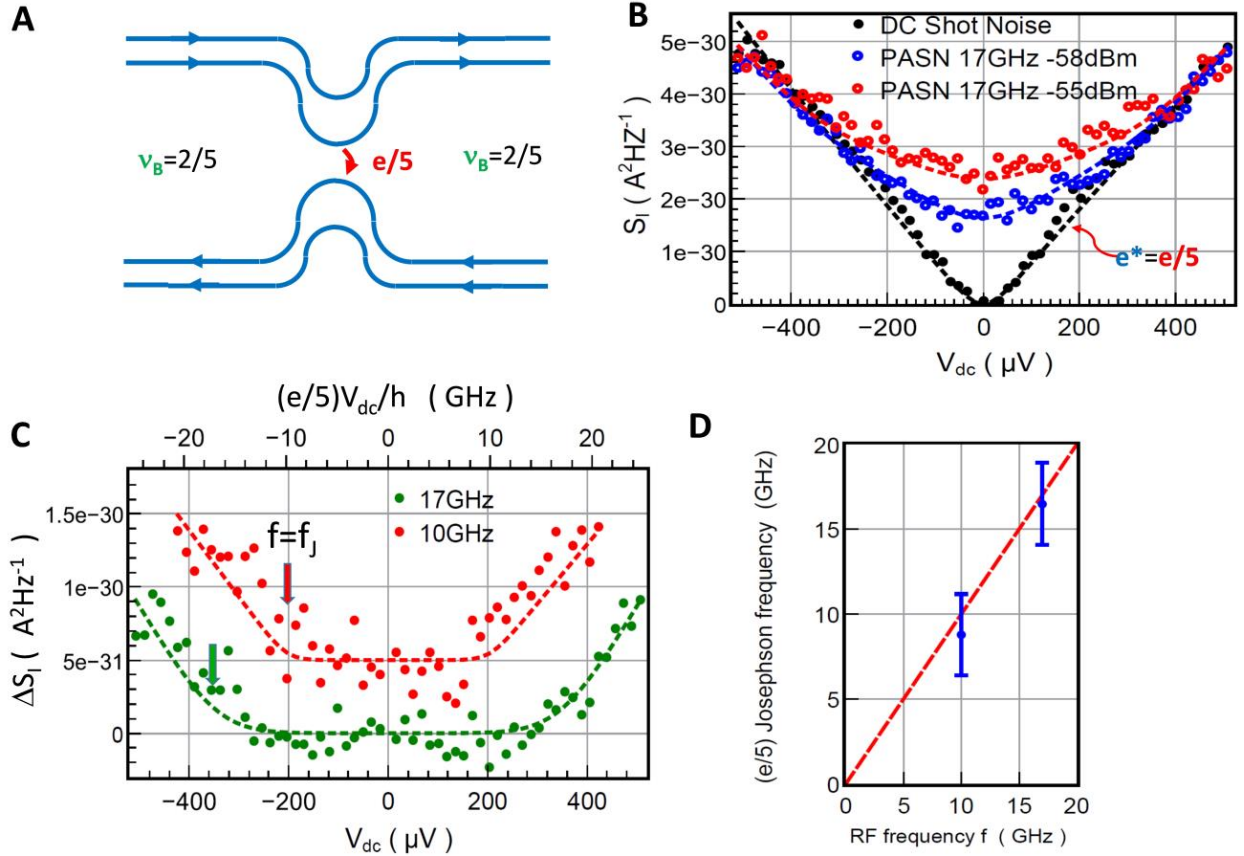

**Fig. 3. Josephson relation for the 2/5 FQHE state.** (**A**) Chiral edge schematics: the 2/5 inner edge state is weakly reflected, see point (B) of Fig.1B. Here backscattered e*=e/5 carriers contribute to current $I_B$ and shot noise $S_I$. (**B**) Raw shot noise measurements: Black dots show the DC shot noise (i.e. no RF) measured during the 17GHz PASN measurement run. The black dashed line is Eq. 2 using R=0.064 (point (B) of Fig.1A ). Blue and red open circles is the PASN for 17GHz -58dBm and -51dBm RF irradiation. Blue and red dashed line curve are plotting Eq. 1 with $f_J$=(e/5)$V_{dc}$/h (**C**) Excess PASN $\Delta S_I$. Green and red dots correspond to 17 and 10 GHz, respectively. Green and red dashed lines are computed from Eq. 3 using $f_J$=(e/5)$V_{dc}$/h. For clarity, for each curve the corresponding $\Delta S_I$( $V_{dc}$=0) has been subtracted from the excess PASN. (**D**) Determination of e*: a best fit of $\Delta S_I$ gives, for each frequency, the threshold voltages $V_J$=hf/e* above which $\Delta S_I$ rises. They are plotted in units of (e/5)$V_J$/h versus f (blue points with s.e.m. errors

bars); a linear fit gives e*=e/(5.17 ± 0.31 ). For comparison, the red dashed line corresponds to e*=e/5 exactly.

# Supplementary Materials for

## A Josephson relation for fractionally charged anyons.

M. Kapfer, P. Roulleau, M. Santin, I. Farrer, D. A. Ritchie, and D. C. Glattli.

correspondence to: christian.glattli@cea.fr

**This PDF file includes:**

Materials and Methods
SupplementaryText
Figs. S1 to S8

**Other Suplementary Materials for this manuscript includes the following:**

Databases for figures 2, 3, S2-S7 as single zipped archive:
(available under DOI 10.5281/zenodo.2388813 )

**Materials and Methods**

<u>Sample characteristics and fabrication:</u>
The samples are 2DES with electrons confined at the interface of high mobility epitaxially grown GaAs/Al$_x$Ga$_{1-x}$As heterojunctions at 90 nm below the surface. The 90nm depth corresponds to a 40nm undoped AlGaAs buffer layer, with x=0.30 Al content, followed by 40nm AlGaAs Si-doped region and ending with a 10nm GaAs cap layer. The low temperature zero field mobility is 2.0×10$^6$cm$^2$s$^{-1}$V$^{-1}$ and the electron density n$_s$=1,11×10$^{15}$m$^{-2}$. For this density, the bulk filling factor ν$_B$=2/5 corresponds to a magnetic field of ≈11.2 Tesla. Ohmic contacts are realized by evaporating 125 nm Au, 60 nm Ge, 4 nm Ni followed by annealing at 470°C. A shallow mesa etching (H3PO4 phosphoric acid, time 4 minutes) defines the sample. The QPC split gates are realized by e-beam lithography, see Fig.S1 for a SEM image of the sample used. The gate separation is 300nm.

<u>Measurement set-up:</u> an ultra-low temperature cryo-free dilution refrigerator from CryoConcept$^R$ is used to provide a 20 mK base temperature as in (*12*). It is equipped with a dry superconducting coil able to reach 14.5 Tesla. Ultra-low-loss dc-40GHz microwave cables, same as in ref. (*12*), bring the room temperature microwave excitation from an Agilent N5183A RF source to a PCB. The nominal RF power given in the main text is estimated from the RF source power and the fixed 60dB cold attenuators. A nominal -60dBm rf power corresponds to V$_{ac}$$^{nom}$ ≈ 450 μV. This value is expected in the low frequency limit. However the actual V$_{ac}$ can be smaller at frequency above a few GHz due to extra skin depth losses in the coaxial lines. *Dynamical screening* in the FQHE sample may also affects the effective V$_{ac}$. Coplanar waveguides designed by CST microwave Studio$^R$ etched on the PCB bring the radiofrequency to the ohmic contact (0) of the sample, see Fig.1(A) and Fig.S1. The noise measurements are obtained by separately converting the transmitted and reflected current fluctuations into voltage fluctuations at contact (1) and (2) respectively in parallel to a R-L-C resonant circuit with 2,5 MHz resonant frequency and 700kHz bandwidth 700 kHz, with R=20kOhms. The voltage fluctuations are amplified by two home-made cryogenic amplifiers with 0.22 nV/Hz$^{1/2}$ input noise at low temperature, followed by low noise room temperature amplifiers. The amplified fluctuations are passed through Chebyshev filters and then sent to a fast 20Ms/s digital acquisition card (ADLink 9826) inserted in a PC which provides real-time computation of the cross-correlation spectrum. Absolute Noise calibration is done by recording the equilibrium Johnson Nyquist noise when varying the temperature from 20mK to 200mK. Differential Conductance measurements giving the transmission and reflection are made by applying a low 0,7 kHz frequency sub-μV amplitude voltage to contact (0) and sending the amplified AC voltage from contacts (1) and (2) to two Lock-in amplifiers. The measurement accuracy is mostly limited by the large 1/f noise of the cryogenic HEMT amplifier (white noise cross-over at ≈1MHz). The shot noise accuracy is limited by the input white noise of the amplifier and time averaging. For ν$_B$=2/5, the 20kOhm resistor and the ≈5kOhm inductance parallel resistance in parallel with the bulk Hall resistance converts the input noise of 220pV/Hz$^{1/2}$ into 2×10$^{-27}$A$^2$/Hz equivalent current noise power. Using cross-correlation and noise averaging during the measurement time τ=300s with ≈350kHz effective detection bandwidth, the accuracy of a raw noise data point is ~3×10$^{-31}$A$^2$/Hz.

<u>Data analysis:</u> In the IQHE regime the finite temperature expression for cross-correlated shot noise of a single channel is:



$$S_I^{dc}(V_{dc}) = 2eR(1-R)\frac{e^2}{h}V_{dc}(\coth(eV_{dc}/2k_BT_e) - 2k_BT_e/eV_{dc}) \tag{S1}$$

where R is the reflection coefficient and $T_e$ the electronic temperature. In the FQHE regime, for very weak reflection coefficient, the backscattering current $I_B(V_{dc})$ is almost linear with voltage, see Fig.S2(a) and S3, and one can use the following expression :

$$S_I^{dc}(V_{dc}) = 2e^*R(1-R)gV_{dc}(\coth(e^*V_{dc}/2k_BT_e) - 2k_BT_e/e^*V_{dc}) \tag{S2}$$

For the very small R used in this work, Eq. (5) identifies to Eq.(2) but written differently, g being the conductance of the open fractional channel considered. To find the excess PASN $\Delta S_I$ given by Eq.(3) we subtract the appropriate amount of DC shot noise by controlling the weighted $|p_0|^2$ so as to obtain a constant noise at low bias voltage in the range $|V_{dc}|<hf/e$. This provides a measure of $|p_0|^2=J_0(\alpha)^2$. We then observe that the noise is flat over a larger voltage scale given by $|V_{dc}|<(hf-k_BT_e)/e^*$. However there is a small $T_e$ increase, several tens of mK, due to RF heating and, when subtracting the DC shot noise at the lowest temperature, this prevents a perfect cancellation of noise variation at small bias in the range $|V_{dc}|<k_BT_e/e^*$, see figures S4(d) and (e). We then subtract from the PASN data the DC shot noise given by (S2) and adjust $T_e$ for the best noise variation cancellation over the full range $|V_{dc}|<hf/e$. We note that only the form of PASN given by Eq.(1), with a discrete sum of shifted voltages, can lead to a flat noise variation. If the PASN was given by a trivial classical time averaging of the noise $<S_I^{DC}(V(t))>$ with $V(t)=V_{dc}+V_{ac}\cos(2\pi ft)$, such cancellation would be impossible, see discussion below and Fig.S7. Importantly, the low voltage bias constant noise signalling the cancellation of the DC shot noise, occurs because the noise is symmetric and linear with respect to bias voltage $V_{dc}$.

**Supplementary Text**

Current and DC shot noise in the WB regime:

The Luttinger Liquid (LL) approach to the FQHE chiral edges predicts a non-linear dependence of the backscattered current with voltage $I_B(V_{dc})$. In experiments, LL predictions are qualitatively observed but a clear agreement is lacking as the theory can't take into account all real experimental details such as: - the long range Coulomb interaction necessary to describe the LL bosonic modes (edge magneto-plasmons); - the energy dependence of the coupling in the WB regime;- the non-local coupling of the QPC; - the local edge channel reconstruction at the QPC.

Fig.S2(a) shows $I_B(V_{dc})$ for point (B) ( Vg=-0.03V) of Fig.1(B) main text as observed during the 10GHz PASN experimental runs. Here, a small drift of the QPC transmission gave a measured R=0.051 slightly different with the R=0.064 measured during the 17GHz PASN experimental runs. We observe an almost linear backscattered current variation. Fig.S2(b) shows DC shot noise data (black dots), a comparison with Eq.(2) (red dots) computed using the measured $I_B$, e*=e/5, and $T_e$=20mK. The blue line curve is also Eq.(2) but using a linear $I_B$ with constant R=0.051.

Fig.S3 similarly shows $I_B(V_{dc})$ for the WB regime of the 1/3 FQHE state corresponding to point (A) (Vg=-0.090V) of Fig.1(B) main text and to the conditions of the DC shot noise



measurements displayed in Fig.2(B) main text. The current is again found almost linear at large bias and shows a steeper slope at low bias in qualitative agreement with the LL predictions. The average reflection coefficient is R=0.026. A constant R gives a good representation of the DC shot noise as displayed in Fig.2(B) with e*=e/3.

Excess PASN for e*=e/3: In this section we show some intermediate data used to calculate the average excess PASN displayed in figure 2(C).

Fig.S4(a-c) shows the excess PASN measured in the WB regime of the 1/3 FQHE state for -61, 63 and -67dBm RF power at 22GHz. (a) and (c) correspond to the raw PASN data shown in Fig.2(B) main text. Averaging the three excess PASN curves (a), (b) and (c) data gives the blue dot data of Fig.2(C) main text. The parameter $\alpha$ in each figure corresponds to $p_0=J_0(\alpha)$, with $J_n(x)$ an integer Bessel function. The red dashed lines in Fig.S4 are a comparison with data obtained by fixing e*=e/3 (Josephson frequency $f_J=(e/3)eV_{dc}/h$ ) and using $p_1=J_1(\alpha)$. Here, $f_J=f$ for $V_{dc}=270.6$ µV. We observe that, within experimental accuracy, the amplitude of the excess PASN variation is well taken into account by $|p_1|^2$ deduced from $|p_0|^2$. Note that $|p_1|^2 \approx (1-|p_0|^2)/2$ as, according to the small $\alpha_s$ found, two-photon emission-absorption processes are negligible and $|p_2|^2 \approx 0$.

In Fig.S4(a-c) a slightly larger electronic temperature, due to RF heating, has been introduced. This corresponds to a small rf heating increasing with power. To be consistent, finding the right $p_0$ which gives a flat Excess noise variation at low voltage requires to subtract to the full PASN data a DC shot noise including the actual electron temperature. Indeed, the subtraction of the DC noise weighted by $|p_0|^2$ does not fully cancels the noise variations at very low DC voltage ( $|V_{dc}|<k_BT_e/e^*$ ) if only the base temperature is used and a small positive cusp is observed as simulated in Fig.S4(d). To remedy this, instead of using the raw low temperature DC shot noise data, we use for the subtraction the theoretical DC shot noise in which we have let the electronic temperature $T_e$ free to obtain a constant excess PASN around zero voltage. The optimum $T_e$ gives an estimation of heating effects. In practice, the condition for finding the flattest noise variation in the range $|V_{dc}|<(hf-k_BT_e)/e^*$ is obtained by minimizing the excess noise variance while varying $p_0$ and $T_e$. The estimated temperature $T_e$ for each RF power is shown in the legend of Fig.S4(a-c) and Fig.S4(f) shows how $T_e$ varies with power. The temperature accuracy is ±10mK and $\alpha$ is known within 20% accuracy due to the noisy PASN data.

In the same WB regime of the local 1/3-FQHE state formed at the QPC we show in Fig.S5(b and c) data for f=10GHz and for the two RF power -70dBm and -65dBm used to calculate the 10 GHz average excess PASN displayed in Fig.2(C) main text. Fig.S5(a) shows in black the DC shot noise data (dots) and a DC shot noise fit (dashed line) using e*=e/3, $T_e$=20mK and R=0.026. The blue and green open circles are respectively the PASN data for a 10GHz -70dBm and -65dBm RF power. The dashed lines are full PASN calculation using $\alpha$=0.8 and 1.17 respectively. As previously done, $\alpha$ is deduced from the $|p_0|^2$ giving a constant variation in the excess PASN noise as shown in Fig.S5(b) and (c). The red dashed line are not fit but comparison with Eq.(2) using $|p_1|^2=(1-|p_0|^2)/2$, assuming negligible higher order photon processes and using $f_J=(e/3)V_{dc}/h$. Here $f_J=f$ for $V_{dc}=123$ µV. Again, the deduced $|p_1|^2$ accounts remarkably well for the amplitude of the excess PASN variation observed for each RF power.

Excess PASN for e*=e/5: Finally, we show some extended data for the weak backscattering regime of the 2/5 inner edge channel characterized by e*=e/5 charge carrier for $V_G$=-0.03 V (point



(B) of Fig.1(B) main text). Fig. S6(a-c) shows the three excess PASN data for power -60, -58 and -55 dBm used to generate the average excess PASN data displayed in Fig.3(C) of the main text. Here the weaker noise due to the weaker one fifth charge and the smaller backscattered current make the data more affected by the detection noise. The accuracy of the deduced α is ≈30% and the accuracy of the deduced temperature resulting from RF heating is ±15mK.

Simple justification of Eq.(1) and Josephson terminology: Eq.(1) has been derived in a general frame (*18*) and is implicitly found in some PASN equations derived in the FQHE regime in refs. (*17,19,20,34-37*). Here we provide the reader with a simple physical derivation, trying to avoid all cumbersome quantum formalism. In this heuristic derivation we assume well defined quasiparticles of charge e* as usually considered in FQHE tunneling models.

We consider two very weakly (tunnel) coupled reservoirs (L) and (R) at electrochemical potential $\mu_L$ and $\mu_R$ whose ground state many-body wave-function is denoted $|\mu_L\rangle$ and $|\mu_R\rangle$ respectively and the tensor product is denoted $|\mu_R\rangle|\mu_L\rangle$. An observable $\tilde{A}$, has an expectation value:

$$A^{DC} = \langle\mu_L|\langle\mu_R| \tilde{A} |\mu_R\rangle|\mu_L\rangle \tag{S3}$$

$A^{DC}$ can be the shot noise considered here or another DC transport property in a non-equilibrium situation when $\mu_L-\mu_R=e^*V_{dc}$ and e* is the charge of the carriers experiencing the static voltage $V_{dc}$.

We now consider an AC voltage with amplitude $V_{ac}$ and frequency *f* added to the left reservoir potential while the right electrochemical potential is fixed. Assuming for simplicity a uniform $V_{ac}$ in the left reservoir, all left carriers will experience the time dependent potential: $e^*V_{ac}(t)=e^*V_{ac}\cos(2\pi ft)$.

The phase $e^{-i2\pi\varepsilon t/h}$ of a carrier with energy $\varepsilon<\mu_L$ contributing to state $|\mu_L\rangle$ will acquire an extra time dependent phase $\phi(t)=e^*V_{ac}\sin(2\pi ft)/hf$. Defining the Floquet amplitudes:

$$p_l = \frac{1}{2\pi}\int_t^{t+T}\exp(i2l\pi ft')e^{i\phi(t')}dt' \tag{S4}$$

we see that all carriers experiencing the AC potential are dynamically scattered into energies $\varepsilon \to \varepsilon + lhf$ with probability amplitude $p_l$. As a consequence, the whole left ground state ends in a superposition of states with the extra (factorized) phase $e^{il2\pi ft}$. The expectation value of $\tilde{A}$ becomes:

$$A = \left(\sum_{l'}p_{l'}e^{il'2\pi ft}\langle\mu_L|\right)\langle\mu_R|\tilde{A}|\mu_R\rangle\left(\sum_l p_l e^{-il2\pi ft}\right)|\mu_L\rangle \tag{S5}$$

In the experiment, one measures the time average value of A:

$$\overline{A} = \sum_l |p_l|^2 \left(e^{il2\pi ft}\langle\mu_L|\right)\langle\mu_R|\tilde{A}|\mu_R\rangle\left(e^{-il2\pi ft}|\mu_L\rangle\right)$$
$$= \sum_l |p_l|^2 \langle\mu_L+lhf|\langle\mu_R|\tilde{A}|\mu_R\rangle|\mu_L+lhf\rangle \tag{S6}$$

where, in the second expression, the time dependent phase has been absorbed into a global shift of the electrochemical potential. For electrochemical potential difference $\mu_L-\mu_R=e^*V_{dc}$ we see from Eq. S3 that:



$$\overline{A} = \sum_l |p_l|^2 A^{DC}(e^*V_{dc} + lhf) \quad (S7)$$

Eq. S7 was derived without specifying the nature of the ground state, except the existence of quasiparticles at relevant energies near the Fermi energy. It leads to Eq.(1), main text, for shot noise. Eq. S7 applies to transport properties such as current (*15,18,34*), noise (*17-20*) or even thermal current (*34-36*) and can be found implicitly or explicitly written in the form of Eq. S7 in the quoted references. For the current, looking for experimental signatures of Eq S7 is difficult because of the lack of clear marked singularities of the DC current. Although Luttinger liquid (LL) models aiming at describing transport in FQHE edges predict zero bias singularities (*15*), they are found weaker than expected in the very WB regime (see above). This prevents for a convincing observation of photon assisted effects using current. Shot noise however appears more appropriate as it always shows a zero bias voltage singularity. The shot noise provides at the same time a determination of the carrier charge e* based on two distinct physical mechanisms: -1) the charge of the carriers able to perform elementary photo-excitation transitions (PASN) as demonstrated in the present work via the Josephson relation; -2) the charge granularity (*4-6,9,10*) via the DC shot noise given by Eq.(2).

To conclude this section, we note that the form of Eq. S7 leads to a robust comparison of voltage to frequency via the Josephson relation f=e*$V_{dc}$/h. The only ingredient for such a relation to exist is a periodic modulation of the quasiparticle phase leading to finite Floquet amplitude probabilities. The exact value of the $p_l$ is not relevant as it will not affect the Josephson relation. Due to *dynamical screening effects*, the assumption of uniform $V_{ac}$ amplitude in the left reservoir may be not correct and the actual $p_l$ values may be different from $p_l = J_l(e^*V_{ac}/hf)$. This issue will deserve further studies.

Regarding the Josephson terminology we think the expression "Josephson relation" is appropriate as discussed in the main text introduction. The expression "Josephson frequency" has been widely used in theoretical papers dealing with the tunneling of fractional charges in FQHE, see (*15-18*) and the non-exhaustive list of the extra references (*38-45*)

<u>Photon-assisted process versus classical AC voltage averaging:</u> The experimental observation of the excess PASN is possible because of the discrete nature of the photon assisted process leading to Eq.(2), main text. The PASN being a discrete sum of DC shot noise curves with voltage shifted by lhf/e* and weighted by $|p_l|^2$ it is then possible to cancel the l=0 term. For symmetric and linear DC shot noise variation, when hf>>$k_BT_e$ this leads to the characteristic flat variation of the excess PASN for $V_{dc}$<hf/e*. This become clearer if we write Eq.(3) as:

$$\Delta S_I(V_{dc}) \approx \Delta S_I(0) + 4g_{1(2)}R\langle |p_1|^2\rangle \left[ \frac{hf - e^*V_{dc}}{e^{\frac{hf-e^*V_{dc}}{k_BT_e}} - 1} + \frac{hf + e^*V_{dc}}{e^{\frac{hf+e^*V_{dc}}{k_BT_e}} - 1} \right] \quad (S8)$$

Here $g_2$=(2/5-1/3)$e^2$/h and $g_1$=(1/3)$e^2$/h as defined in the main text.

In this section, we show that a trivial classical adiabatic averaging of the DC shot noise obtained by replacing $V_{dc}$ by V(t)=$V_{dc}$+$V_{ac}$cos(2πft) and taking the time average of the noise is not able to give a flat noise variation at low bias by subtracting any amount of DC shot noise from the PASN data. Fig.S7(a) shows a simulation of the DC shot noise for $T_e$=30mK and e*=e/3 and of the PASN for α=0.60 and f=22GHz, black and red curves respectively. For comparison, the green curve shows the classical adiabatic noise averaging with same AC amplitude $V_{ac}$=αhf/e*. We observe a wide parabolic variation of the adiabatic noise around zero voltage contrasting with the sharp zero bias singularity of the DC shot noise. It is unlikely that any amount of DC shot noise subtracted from the classically averaged noise would provide a flat variation over the voltage range $|V_{dc}|<hf/e^*$. Indeed, Fig.S7(b) shows, in red, the (quantum) excess PASN using $\alpha=0.6$ ($|p_0|^2=0.8317$) and, in green, the DC shot noise subtraction from the adiabatic noise with same weight $|p_0|^2$. The flat low bias variation of the excess PASN contrasts with the wavy variation of excess adiabatic noise.

Perspectives: a time resolved source of FQHE anyons based on levitons: Observing PASN in the FQHE regime is an important step toward the realization of time-resolved single anyon sources based on periodic voltage pulses. With such source, we expect that single abelian or non-abelian anyons can be emitted in a FQHE edge channel in a similar way photons can be emitted using single photons sources. Not claiming this can be directly useful for adiabatic quantum computing, we think that the time domain operation opens a realm of new quantum interference involving anyons while conventional interferometry experiments using DC voltage sources and aimed at performing anyon interferometry (47,48) are debated (49-51). Combining two sources and the relative time delay between the emission of two anyons in the input of a beam splitter (a Quantum Point Contact) will allow to perform Hong Ou Mandel experiments with anyons providing a direct measurement (13,46) of the statistical anyon statistics angle $\theta_S$ ($\theta_S=\pi/3$ or $\pi/5$ for abelian FQHE anyons with charge $e^*=e/3$ or $e/5$ respectively). Indeed, let us consider two anyon simultaneously arriving at the separate left and right inputs of a beam splitter whose transmission amplitude is t and reflection amplitude is ir and $|t^2|+|r^2|=1$ for unitary scattering matrix. The two-particle coincidence amplitude probability b(1,2) to find particle (1) at the right output and particle (2) at the left output is $b(1,2) = t^2 a(1,2) + (ir)^2 a(2,1)$ where a(1,2) is the amplitude of having particle (1) coming from the left input and (2) on the right input and a(2,1) is particle (2) coming fom the left and particle (1) coming from the right. The two input case correspond to swaping (braiding) (1) and (2) and so $a(2,1)=e^{i\theta_s} a(1,2)$. For half transmission one gets the coincidence probability $P(1,2)=|b(1,2)|^2=1/2(1-\cos(\theta_s))$. This corresponds to bunching for Bosons ($\theta_s=0$) and Pauli exclusion for Fermions ($\theta_s=\pi$) and something in between for anyons. Including the overlap factor $g_2(\tau)$ (or second order coherence in quantum optics language ) for incoming particles time shifted by $\tau$, we get: $P(1,2,\tau)=1/2(1-g_2(\tau) \cos(\theta_s))$. As $g_2(0)=1$ and $g_2(\infty)=0$, the time resolution provides a safe way to measure $\cos(\theta_s)$ and so demonstrate anyon statistics, see Eq. (13) in Ref. *(53)*.

Knowing the coincidence property, is it easy to derive the cross-correlated noise is ~ $(1+g_2(\tau)) \cos(\theta_s)$.

Similarly letting interfere three or more non-abelian anyons from the FQHE state with $\nu_{Bulk}=5/2$ would give access to *non-abelian* statistics interferences by multi-particle coincidence measurements

Leviton source in the FQHE:
A Leviton is a single charge pulse generated by applying a Lorentzian voltage pulse on a contact (*12*). It is made from electron like excitations only or hole-like excitations only but does not contain any mixture of electron and hole excitations. This is a minimal excitation state (*33*) which has the property to generate a minimal noise when partitioned by an electron beam-splitter like a Quantum Point Contact.

Let us consider periodic voltage pulses V(t) on the contact of a FQHE edge with for example $\nu_B$=2/5. The pulses create periodic current pulses I(t)=V(t) $e^2$/5h injected in each of the two 2/5 FQHE edge channels. The FQHE edge channel carriers acquire a periodic time-dependent phase $\phi(t) = \frac{e^*}{\hbar} \int_{-\infty}^{t} V(t')dt'$. Let $p_l$ be the Floquet amplitudes as defined previously for a harmonic excitation. A pure electron (hole) like minimal excitation, a Leviton, can be created if and only if all the $p_l$ are zero for $l<0$ (or $l>0$), i.e. the Fermi sea is shifted only to positive (or negative) energies. To achieve such property it is necessary that: 1) the phase increment for each periodic pulse is 2π and 2) each voltage pulse shape is Lorentzian in time. Condition 1) shows that *only an integer* charge $Q = \int_{t}^{t+T} I(t')dt' = e$ can be injected per period T=1/f in each of the FQHE edge channels (*20*). Trying to inject a non-integer charge would create a complex mixture of quasi-electron and quasi-hole FQHE excitations, see the dynamical orthogonality catastrophe discussed in (*34*). Condition 2) says that only a special pulse shape would work. As shown in Ref. (*12,20,34*) sine-wave pulses $V(t) = V_{ac}[1 - \cos(2\pi ft)]$ even injecting an integer charge e (tuning the amplitude to $V_{ac} = hf/e^*$ or: α=1) are not appropriate to generate levitons. In the present experiment such integer charge sine wave pulse have been incidentally generated in Fig.S3(b) at 17GHz where α≈1 for $V_{dc}$=hf/(e/5)≈350 µV.

Poissonian source of time resolved anyons
Once the periodic charge e Leviton source in FQHE is operational, the next step is to send the levitons towards a QPC in the weak backscattering regime to generate a weak backscattering of fractional anyons of charge e* as done in the present experiment while using sine-wave excitation. The *deterministic* character of the periodic generation of Leviton *will be lost* as the quasiparticles will be created following a Poisson's statistics, but we expect the anyons will inherit the time domain properties of the levitons (*45*). Fig.S8 shows a sketch of the Poissonian stochastic single anyon sources and their use for Hong Ou Mandel correlation for $\nu_B$=1/3. A similar set-up has been theoretically considered in Ref. *(52)* but considering only DC voltage sources. Positive correlations were found but the lack of time control make the result less convincing. Generalization to other FQHE states, including non-abelian, is straightforward.

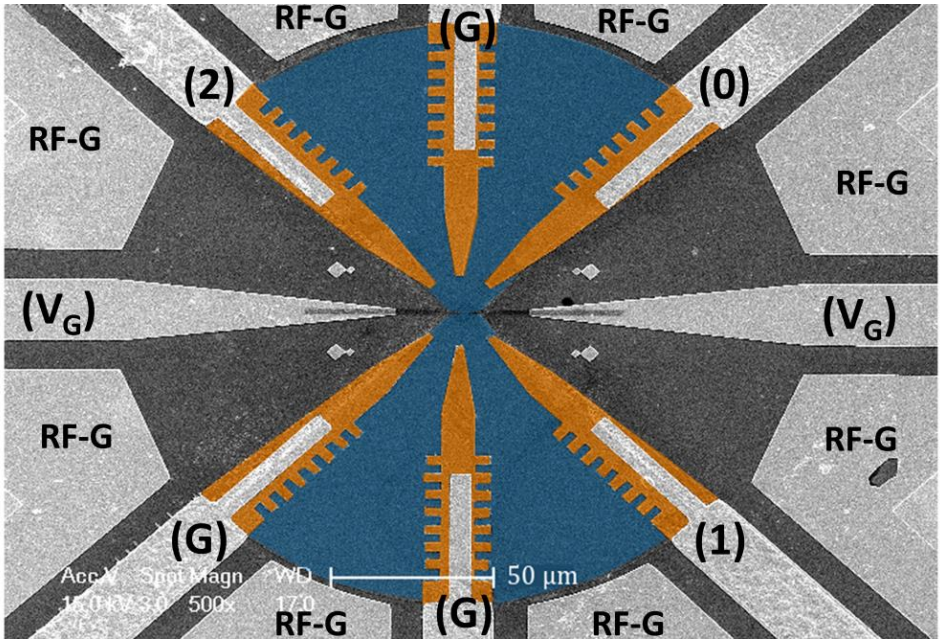

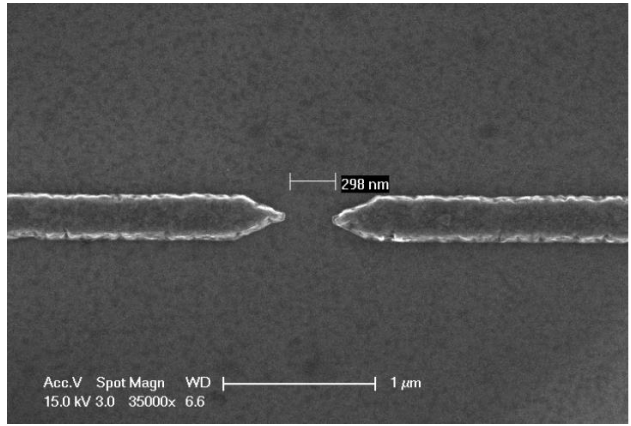

**Fig. S1.**
*Top image*: SEM view of the sample (colorized image). Ohmic contacts are in yellow, the 2DEG in blue. The magnetic field sign is chosen such that the edge channels run clockwise along the 2DEG edge. The white bar in lower right corresponds to a scale of 50μm. Ohmic contact (0) is used to apply both DC and AC voltage, contact (1) to detect the transmitted current and contact (2) the backscattered current. All contacts (G) are grounded. The metallic gates labelled RF-G are all grounded and used to screen the RF electric field. The two RF-G gates on both side of the metallic strip leading to contact (0) provide a coplanar waveguide guiding the RF excitation to contact (0). The indentation of ohmic contacts is used to lower the ohmic contact resistances (typically ~100 Ohms ). The metallic strips labelled $V_G$ and ending into thin dark grey metallic strips in the middle of the sample form the Quantum Point Contact split gate. A detail of the split gate is shown below.
*Bottom image*: SEM view of the QPC split gates. The gate separation is 300nm

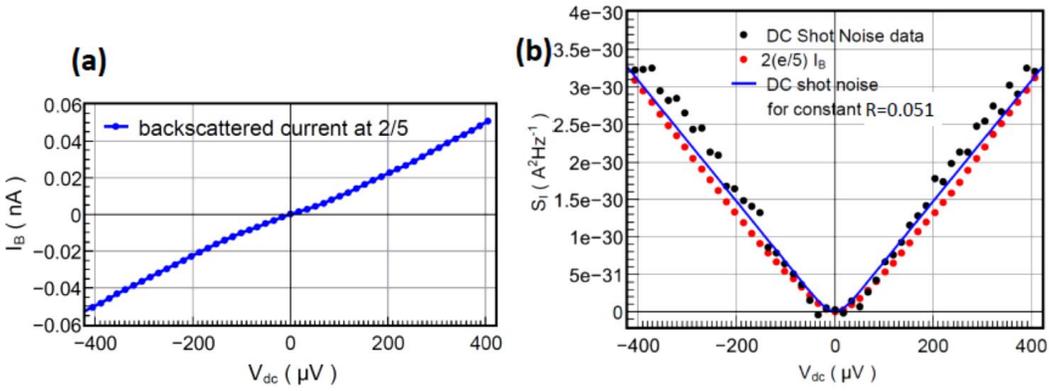

**Fig. S2.**

(a) Backscattered current versus DC voltage fro $V_G$=-0.03V, point (B) of Fig.1(B) as measured during the 10 GHz run. Here R=0.051 is found, different from R=0.064 during the 17GHz run, due to small sample drift.

(b) DC shot noise (black dot); comparison with Eq.(2) (red dot) using $I_B$ measured in Fig.S2(a) and e*=e/5; computed shot noise with constant R (blue solid line) and e*=e/5.

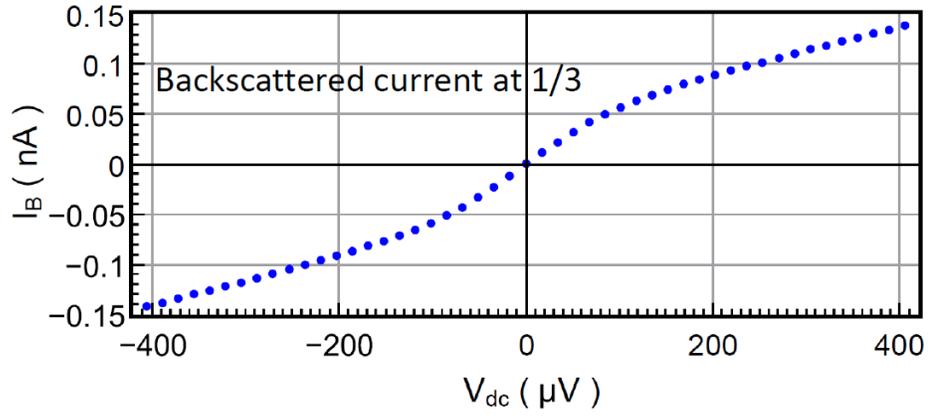

**Fig. S3.**

$I_B(V_{dc})$ for $V_G$=-0.090 V corresponding to point (A) of Fig.1(B), weak backscattering of 1/3 FQHE state. The average R=0.026 gives a good representation of the DC shot noise displayed in Fig.2(B) with e*=e/3

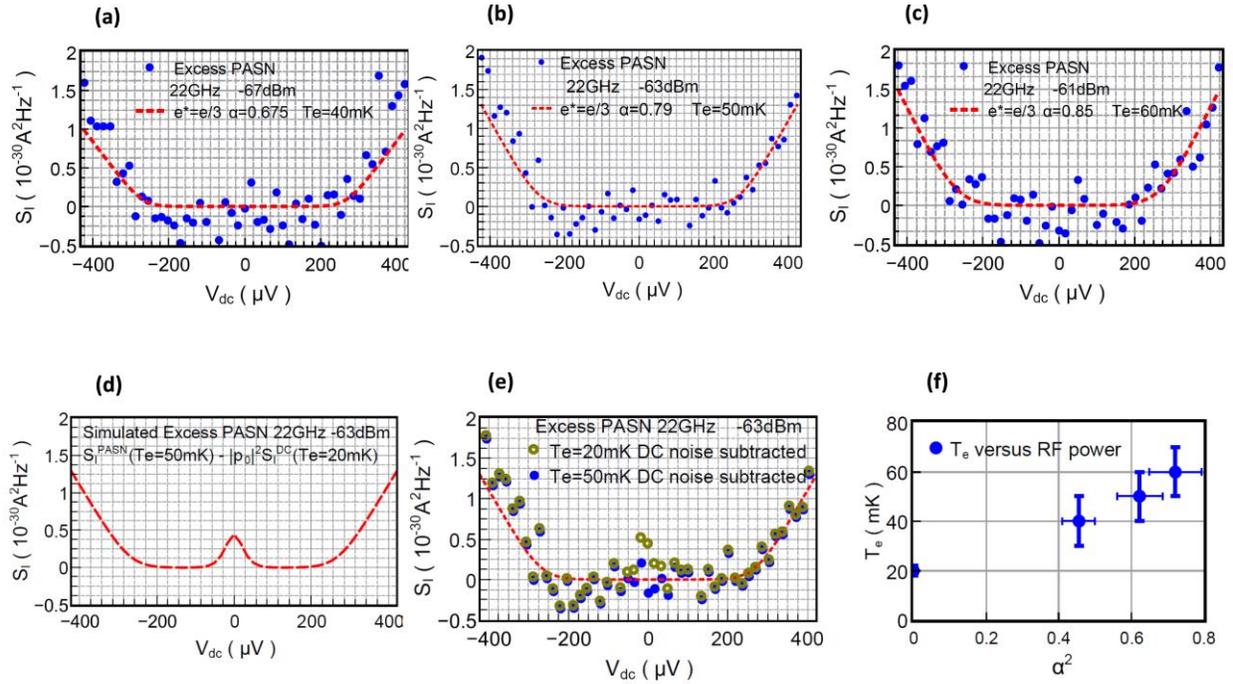

**Fig. S4.**

Excess PASN for 22GHz and (a) -67dBm, (b) -63dBm and (c) -61dBm Rf power. The deduced the parameter $\alpha$ is respectively 0.675, 0.79 and 0.85. Blue dots are excess PASN data and the dashed red curves a comparison with Eq.(3) using $f_J=(e/3)V_{dc}/h$ and the deduced $\alpha$. The average of the three set of data is done to produce the 17GHz average excess PASN in Fig.2(C). For the Excess PASN data shown in (a-c) a temperature $T_e$=40, 50 and 60mk respectively has been used in the subtracting the DC shot noise weighted by $|p_0|^2$. If the base 20mK temperature were used a peak would have been observed near zero voltage in the excess noise as simulated in (d). The excess noise data calculated by subtracted the DC shot noise with base temperature is shown as brown open circle. We observe a peak in the data compared with the blue dots and red dashed line taken from Fig.S4(b) for comparison. In (f) is shown the temperature increase versus $\alpha^2$ ($\propto$ to the measured rf power).

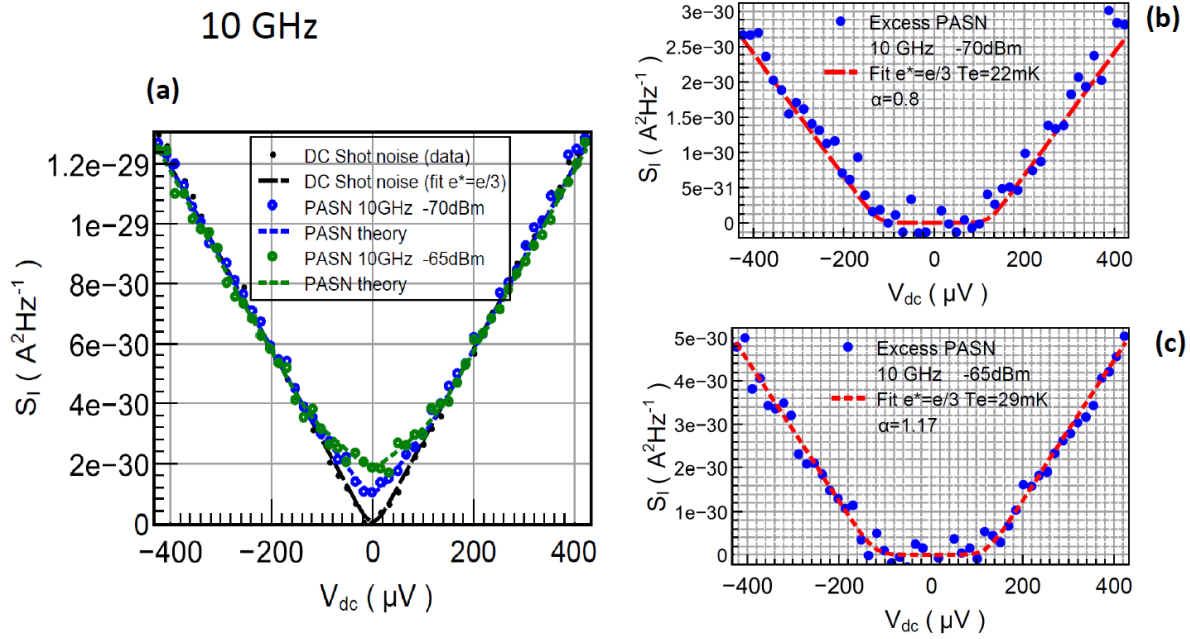

**Fig. S5.**

(a): DC shot noise and (full) PASN at 10 GHz and for -70dBm and -65dBm RF power. (b) and (c): Excess PASN (blue dots) and fit to Eq.(3) (red dashed line) using the deduced $\alpha$ and $f_J=(e/3)V_{dc}/h$ for 870 and -65dBm RF power respectively.

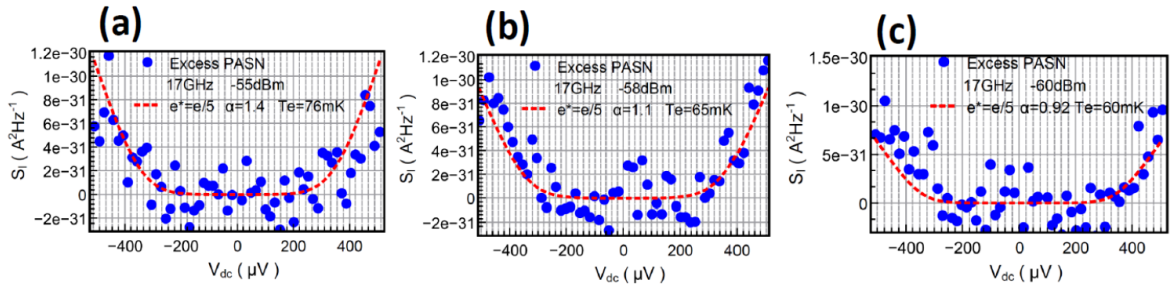

**Fig S6.**

Excess PASN for the weak backscattering regime of the 2/5 inner edge channel characterized by $e^*=e/5$ charge carrier for $V_G=-0.03$ V (point (B) of Fig.1(B)). Blue dots are data and dashed red lines comparison to Eq.(3) using $f_J=(e/5)V_{dc}/h$ and the deduced $\alpha$ for -55 (a), -58 (b) and -60 (c) dBm RF power.

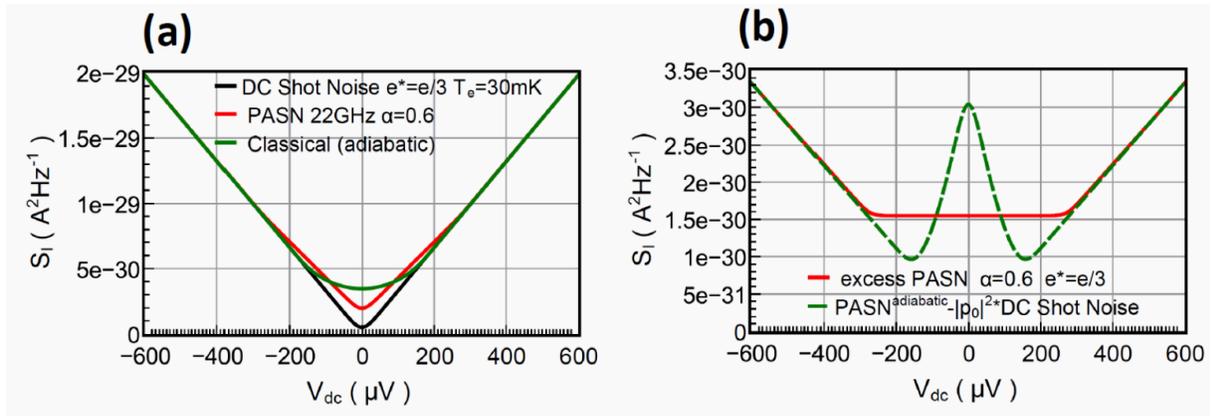

**Fig. S7.**

(a) Red curve: simulated PASN for e*=e/3 and α=0.6; black curve: DC shot noise; green curve simulated adiabatic averaging of DC shot noise.

(b) Simulated Excess PASN for the quantum case (red curve) and the classical adiabatic averaging case (green curve)

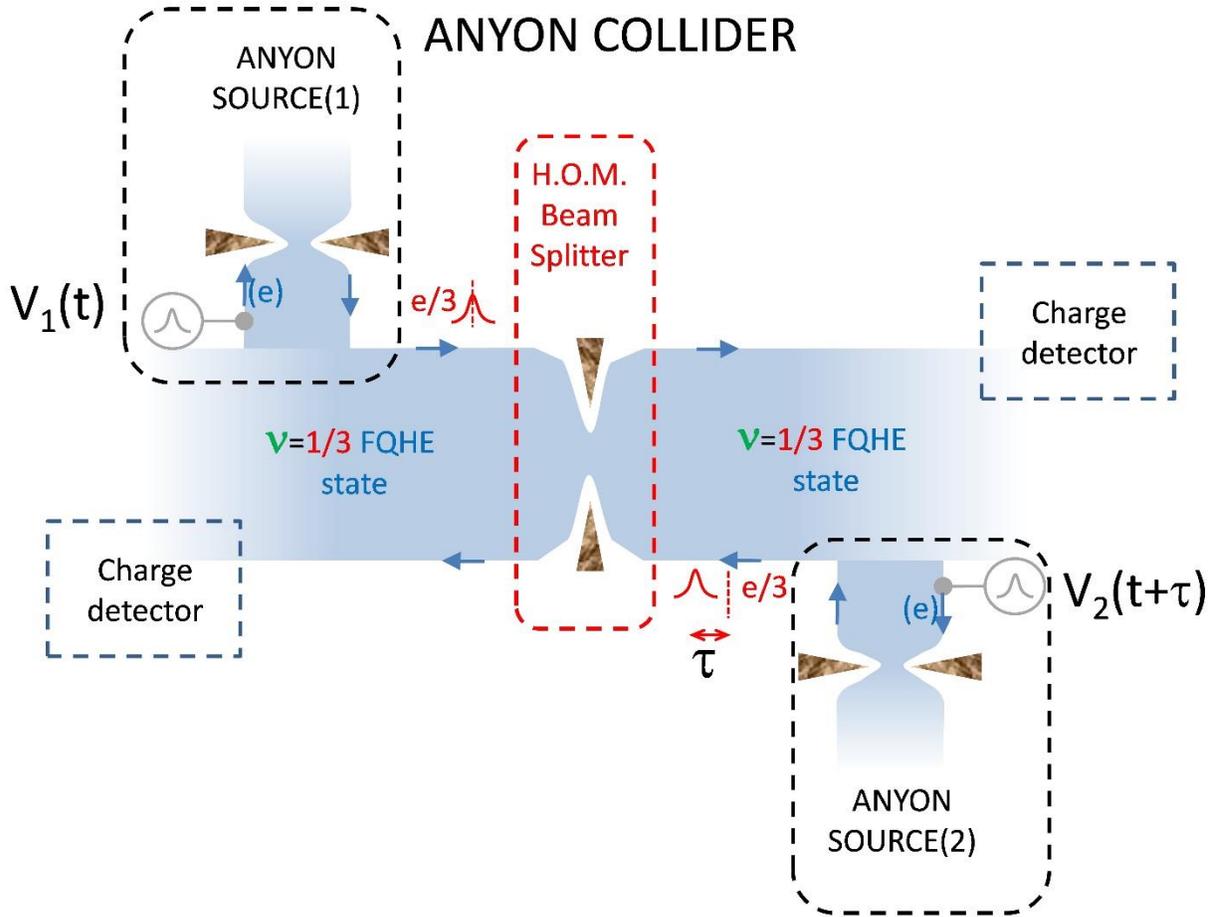

**Fig. S8.**
Hong Ou Mandel correlation experiments aimed at measuring the statistical anyon angle. Two anyon sources are realized by using Lorentzian voltage pulses $V_1(t)$ and $V_2(t)$ to send levitons of charge *e* to a QPC in the WB regime. The levitons generate a Poissonian source of e/3 anyons keeping their time properties. The anyons are then send to the input of a third QPC (in the middle) to be mixed and produce Hong Ou Mandel correlation measured by cross-correlation of charge detection events. The outcome results from the quantum interference of braided and non-braided anyons and thus brings information of the statistical angle. A time delay between the two sources allows for an unambiguous measure of the statistical angle. The present work demonstrate that such anyon sources can be realized with the incremental modification to replace sine-wave pulses by periodic Lorentzian pulses.